# Highly Tunable and Strong Bound Exciton in $MoSi_2N_4$ via Strain Engineering

Dan Liang[1, 2], Shi Xu[3], Pengfei Lu[1*], Yongqing Cai[2*]

[1]*State Key Laboratory of Information Photonics and Optical Communications and School of Electronic Engineering, Beijing University of Posts and Telecommunications, Beijing 100876, China*

[2]*Joint Key Laboratory of the Ministry of Education, Institute of Applied Physics and Materials Engineering, University of Macau, Macao SAR 999078, China*

[3]*Department of Physics and Chemistry, Faculty of Science and Technology, University of Macau, Macao SAR 999078, China*

*E-mail: (P.L.) photon.bupt@gmail.com; (Y.C.) yongqingcai@um.edu.mo

**ABSTRACT** Motivated by the recently synthesized layered material $MoSi_2N_4$, we investigated excitonic response of quasiparticle of monolayer $MoSi_2N_4$ by using $G_0W_0$ and Bethe-Salpeter equation (BSE) calculations. With a dually sandwiched structure consisting of a central $MoN_2$ layer analogue of $2H$-$MoS_2$ capped with silicon-nitrogen (SiN) honeycomb outer layers, $MoSi_2N_4$ possesses frontier orbitals confined at the central $MoN_2$ layer with similar sub-valley at K-point as $2H$-$MoS_2$. The valley splitting (~130 meV) due to the spin-orbital coupling (SOC) gives rise to a doublet in the spectrum. Excitons in $MoSi_2N_4$ shows a strong binding energy up to 0.95 eV with the optical bandgap of 2.44 eV. Both electronic and optical gaps are highly sensitive to tensile strains and become redshift albeit a marginal change of exciton binding energy. With the protection of capped SiN layers, quantum confined excitons in $MoSi_2N_4$ without the need of additional passivation layer like BN would provide a bright new platform for robust emission with partially screened disturbance from environment.

## I. INTRODUCTION

Since the discovery of monolayer graphene [1], tremendous experimental and theoretical efforts for two-dimensional (2D) materials were initiated to explore the exotic physical properties and functionalities [2-4]. Different from the three-dimensional (3D) counterparts, 2D materials exhibit additional pathways for low-energy particles coupling and excitation which allow exotic optoelectronic properties [5]. Owing to the spatial confinement and reduced dielectric screening, Coulomb interactions are significantly enhanced in 2D systems [6]. The quantum confinement effect leads to novel electronic performance including quasiparticles (QP) of electron-hole (*e-h*) pairs and strongly bound excitons [7-9]. In addition, the stacking, sliding and twisting flexibilities in layered structures expand the configurational space which in turn brings about Moiré patterns and collective nature of flat-band excitations [10].

Very recently, through introducing Si to passivate the surface dangling bonds of nonlayered 2D molybdenum nitride ($MoN_2$), a new van der Waals (vdW) layered material $MoSi_2N_4$ in septuple atomic layers of N-Si-N-Mo-N-Si-N was discovered [11]. With a $MoN_2$ layer sandwiched by two Si-N bilayers, monolayer $MoSi_2N_4$ shows an indirect bandgap of ~1.94 eV, a high strength (~ 66 GPa) and a remarkable stability under moisture and ambient conditions. The intrinsic carrier mobilities are predicted to be ~270 $cm^2$ $V^{-1}$ $s^{-1}$ of electrons and ~1200 $cm^2$ $V^{-1}$ $s^{-1}$ of holes, which are substantially higher than that of $MoS_2$ [12]. Since its discovery various theoretical examinations have been performed with respect to point defect [13-14], heterostructures [15-16], and strain effect [17]. As demonstrated by strongly suppressed Fermi level pinning (FLP) and wide-range tunable Schottky barrier height (SBH), owing to a large SBH slope parameter arises from well-protected electronic states by the outlying Si-N sublayers, the $MoSi_2N_4$ is also believed to open up an uncharted territory for the exploration of 2D-material-based device technology [18]. Huang et al. predicted that the $MoSi_2N_4$ MOSFETS have an ultralow subthreshold swing and power-delay product, which have the potential to realize high-speed and low-power consumption devices [19]. $MoSi_2N_4$ also exhibits the potential for electro- and photo-catalytic water-splitting [20-22]. Notably, the $MoSi_2N_4$ has its inner core component of $MoN_2$ resembling 2H transition metal dichalcogenides (TMDs) phase (i. e. 2H-MoS$_2$) while outer capped SiN layer analog of corrugated honeycomb sheet of silicene (half amount of silicon atoms being replaced with nitrogen atoms), a natural question is how does the excitonic response behave in such dually sandwiched structure. Additionally, strain engineering could break the inversion symmetry of the crystal lattice, thus opening a bandgap in its otherwise gapless band structure, like in graphene [23]. Contrary to graphene, TMDs are 2D semiconductors with 0.3 ~ 2.0 eV bandgap, which can be further engineered with strains. The optical gap of monolayer MoS$_2$ is very sensitive to tensile strains, which can be tuned by depositing monolayer MoS$_2$ on different substrates, whereas the exciton binding energy is found to be insensitive to the strain [24]. It is crucial to demonstrate the precise strain dependence of electronic and optical properties.

Here, on the basis of first-principles calculations, we adopted the $G_0W_0$ and Bethe-Salpeter equation (BSE) approaches to examine the QP band structures and optical spectra of monolayer MoSi$_2$N$_4$. Due to the spin-orbital coupling (SOC) effect the valley at valance bands (VB) at the K-point splits which is responsible for the two subpeaks appeared in the lowest-energy absorption peaks. Applying moderate tensile strains induces gradual reduction of indirect bandgap. We found that monolayer MoSi$_2$N$_4$ shows a strong excitonic response which is intrinsically protected by the surface SiN layer. The excitons squeezed at the central MoN$_2$ layer are strongly bound against tensile strains and associated with the absorption in the visible range (Figure 1).

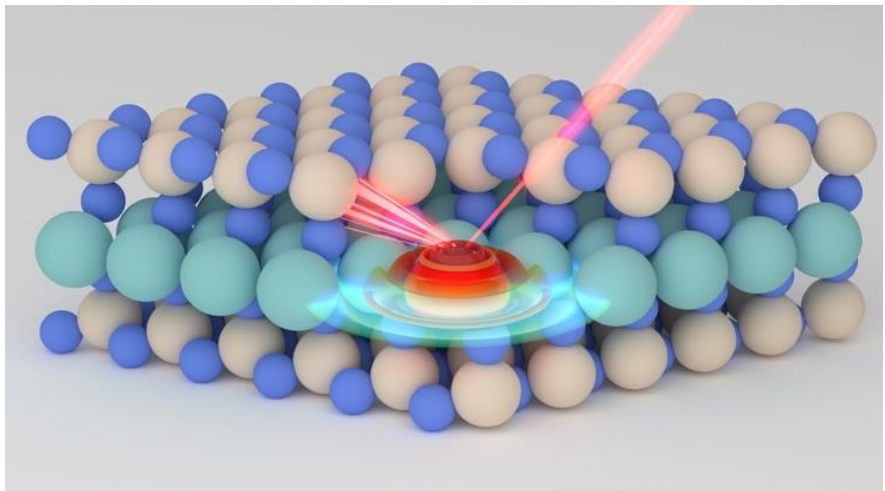

FIG.1. Schematic diagram of MoSi$_2$N$_4$ with strong excitonic response squeezed at the central MoN$_2$ layer.

## II. COMPUTATIONAL DETAILS

All the theoretical calculations were based on the density functional theory (DFT) [25] as implemented in the Vienna *ab-initio* simulation package (VASP) [26-27]. The projector augmented wave method (PAW) [28-29] was performed to represent electron-ion interactions, and the Perdew-Burke-Ernzerhof functional within the generalized gradient approximation (GGA-PBE) [30] was used for the exchange-correlation interactions. The energy cutoff for the plane-wave basis set was chosen to be 500 eV. The structures were relaxed until the forces on each atom were less than 0.01 eV/Å and energy convergence standard was set to $10^{-6}$ eV. A vacuum slab of 15 Å was employed in *z*-direction to prevent interaction between two neighboring surfaces. The *k*-point grid for the first Brillouin zone were sampled on a mesh grid of $9 \times 9 \times 1$. The spin-orbital coupling (SOC) was considered in unstrained $MoSi_2N_4$ for a comparison.

Conventional DFT calculations cannot predict the QP bands, thus the $G_0W_0$ approximation [31-32] that accounts for the many-body electron interactions was adopted to accurately compute the QP bands. More accurate $GW_0$ approach was performed for unstrained $MoSi_2N_4$, where the *G* was iterated four times. The calculated bandgap is 3.23 eV, resulting in only 0.04 eV difference compared to 3.19 eV by using single-short $G_0W_0$. A unified *k*-point mesh $12 \times 12 \times 1$ was used and 300 empty bands were included. Based on the $G_0W_0$ QP bands, the attraction between quasi-electron and hole (on the top of $G_0W_0$ approximation) by solving Bethe-Salpeter equation (BSE) [33] was taken into account. The eight highest valence bands and eight lowest conduction bands were included as basis for the excitonic state. The $G_0W_0$+BSE approach allowed us to obtain the QP bandgaps, optical spectra as well as exciton binding energy.

We explored the reliability of vacuum slab and *k*-point grid density in unstrained $MoSi_2N_4$. Increasing the *k*-point grid density as $9 \times 9 \times 1$, $12 \times 12 \times 1$, $15 \times 15 \times 1$ and $18 \times 18 \times 1$, the calculated indirect bandgaps are 3.25, 3.19, 3.15 and 3.13 eV, respectively. Fixing the *k*-point grid to $12 \times 12 \times 1$, for the vacuum slab of 12, 15, 18 Å, the predicted indirect bandgaps are 3.12, 3.19 and 3.23 eV, respectively. The convergence trend is obvious, despite using the *k*-point grid of $15 \times 15 \times 1$ could achieve a good convergency, while such dense *k*-point grid is computationally time-consuming and accordingly more dense *k*-point grid is needed for the BSE calculation. Similar convergence study was also performed for MoS$_2$ [24], which shows a clear dependence with 12 × 12 × 1 *k*-point grid. Noted that recent theoretical results found the exciton binding energy of 0.95 eV for monolayer MoSi$_2$N$_4$ [34], which is the same as our calculated exciton binding energy using the *k*-point grid of 12 × 12 × 1 and vacuum slab of 15 Å.

**III. DISCUSSION AND RESULTS**

The MoSi$_2$N$_4$ monolayer adopts a non-centrosymmetric hexagonal structure with the *P6m*1 space group (Figure 2a and b). The layer contains seven atomic sublayers of N-Si-N-Mo-N-Si-N along the normal direction with a thickness of 7 Å where the middle Mo sub-lattice locates in the central horizontal mirror plane and sandwiched by two N-Si-N layers. The optimized lattice constants are *a* = *b* = 2.91 Å, agreeing well with reported theoretical [17,22] and experimental results [11]. Monolayer MoSi$_2$N$_4$ exhibits an indirect bandgap of 1.73 eV, where the conduction band minimum (CBM) and valence band maximum (VBM) are located at K- and Γ-point, marked as K$_c$ and Γ$_v$ in Figure 2c respectively. The minimum direct bandgap at K-point, formed between K$_c$ and K$_v$ as shown in Figure 2c, is 2.04 eV. The projected density of states (PDOS) shows that the VBM at K$_v$ is contributed by the hybridization of Mo-*d* and N-*p* states while the CBM at K$_c$ is mainly provided by Mo-*d* states. The partial charge densities of ~2 eV above the Fermi level (Figure 2d) are located at Mo atoms, while the conducting holes nearby the VBM (~2 eV below the Fermi level) (Figure 2e) are mainly located at central Mo and adjacent N atoms, respectively, further demonstrating that the middle N-Mo-N layer contributes to the semiconducting electronic properties of MoSi$_2$N$_4$. Interestingly, such well-protected frontier orbitals are intrinsically protected by the outer SiN layers, without the need of normally adopted additional passivation layers like BN, which is highly desired for robust conduction and optical emission.

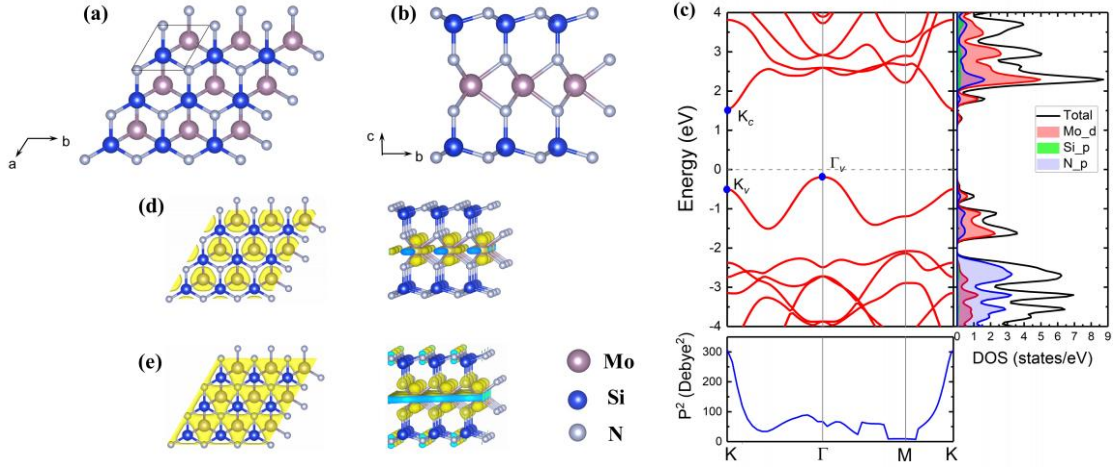

FIG. 2. Top (a) and side (b) views of optimized monolayer MoSi$_2$N$_4$. (c) Band structure, squares of transition dipole moment (P$^2$) from the top band of VB and bottom band of CB, total density of states (DOS) and projected density of states (PDOS) calculated by PBE potential without SOC. The partial charge densities were calculated within 2 eV above (d) and below (e) the Fermi level.

The feature of intimately capped band-edge states prompts us to explore its effect of varying configurational space under strains and examine how the excitonic states are evolved. Due to a more accurate description of many-body electron-electron interactions, the bandgaps predicted by G$_0$W$_0$ are larger over ~1 eV compared to conventional PBE results. Under zero strain, the G$_0$W$_0$ calculation leads to 3.19 and 3.39 eV for the indirect (K$_c$-Γ$_v$) and direct (K$_c$-K$_v$) bandgaps of the QP bands, respectively, as shown in Figure 3a. Next, we introduced biaxial tensile strains on MoSi$_2$N$_4$, from 0 to 6% with a dilation of lattice constants with an increment of 1% per step, to explore the evolution of QP bands by using both G$_0$W$_0$ and PBE approaches. The corresponding lattice constants with strains are given in Table 1. We fully relaxed the strained atomic structure, and found that the symmetries of the lattice and atomic groups maintain, and only the chemical bonds are slightly enlarged due to the applied tensile strains. For the band structures, the CBM moves down and VBM shifts up with increasing tensile strain $\varepsilon$, giving rise to the gradual reduction of the bandgap. Clearly, the PBE and G$_0$W$_0$ results show the same trends, in which both the K$_c$-Γ$_v$ and K$_c$-K$_v$ gaps drop nearly linearly within 0% ~ 6% strain. Notably, the indirect gap between K$_c$-Γ$_v$ decreases from 3.19 eV to 1.89 eV at 6% strain, and the difference between K$_c$-Γ$_v$ and K$_c$-K$_v$ gap increases as increasing strain, which means that moderate biaxial tensile strains are hard to induce an indirect to direct bandgap transition in monolayer MoSi$_2$N$_4$. The exact indirect ($E_{PBE/GW}^{K_c-\Gamma_v}$) and direct ($E_{PBE/GW}^{K_c-K_v}$) bandgap values of strained MoSi$_2$N$_4$ using PBE and G$_0$W$_0$ potential are listed in Table 1, and the strain-gap curves were plotted in Figure 3b. To help to gauge the strains in potential experiment and applications, the coefficients of $dE_g/d\varepsilon$ are fitted and listed as 0.17, 0.11, 0.21, and 0.15 eV for $E_{PBE}^{K_c-\Gamma_v}$, $E_{PBE}^{K_c-K_v}$, $E_{GW}^{K_c-\Gamma_v}$, $E_{GW}^{K_c-K_v}$, respectively. The higher slope for the indirect K$_c$-$\Gamma_v$ gap than the direct K$_c$-K$_v$ gap is associated with a more sensitive movement of the K$_c$ and $\Gamma_v$ states than K$_v$ under strains, leading to different paces of bandgap closure (Figure 3a). Moreover, the degeneracy of valleys in the VBM at K$_v$ and $\Gamma_v$ becomes gradually lifted under increasing tensile strains, which might alter the scattering path of quasiparticles like phonons and electrons/holes.

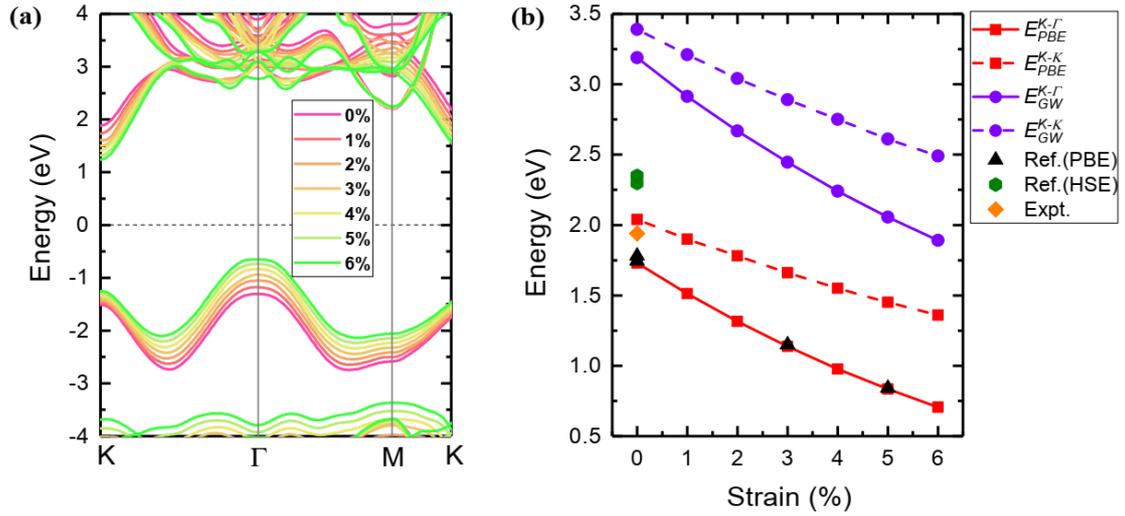

FIG. 3. (a) Quasiparticle band structures of MoSi$_2$N$_4$ under tensile strains from 0% to 6% obtained by G$_0$W$_0$. (b) The variation of indirect (K$_c$-$\Gamma_v$) and direct (K$_c$-K$_v$) bandgaps under tensile strains using PBE and G$_0$W$_0$. Previous theoretical by PBE, HSE and experimental [11,13-14,17] results were also inserted.

Since Mo atom is a transition metal atom, we also considered the SOC in QP band structure calculations as a comparison. The values of K$_c$-$\Gamma_v$ and K$_c$-K$_v$ gaps with SOC are close to that of the gaps without SOC (Table 1). The SOC induces an energy splitting at the VBM at K-point (Figure 4a), leading to a splitting of the gap K$_c$-K$_v$ (3.32, 3.45 eV) of ~130 meV, which is in good agreement with experimentally measured values of ~140 meV [11].

The interaction between quasi-electron and -hole was solved by BSE approach to obtain the optical gap ($E_{GW+BSE}^{opt}$). Comparing the imaginary part of dielectric functions with and without SOC (Figure 4b), there are two intense absorption peaks at ~2.44 and 3.15 eV below the minimum direct $G_0W_0$ bandgap ($K_c$-$K_v$), two subpeaks at ~ 2.42 and 2.53 eV are found with SOC. We observe good agreement with experiments for the first two excitonic peaks in the visible range [11], in which there are a strong peak at ~320 nm (~3.88 eV) and a broad peak at 500 to 600 nm (fitted with two subpeaks at 2.21 and 2.35 eV). The latter two subpeaks observed in experiment correspond to the two direct transitions arising from the splitting of the VB at the K-point. It should be noted that only direct excitations are considered in DFT calculations, while in reality the indirect nature of $MoSi_2N_4$ allows higher energy excitonic states at room temperature via electron-phonon coupling [5]. The less deeply bound excitonic emission associated with those indirect excitons, not considered here, may increase the intensity and/or broaden the observed secondary absorption peaks. The binding energy of exciton is estimated by taking the difference between the minimum direct $G_0W_0$ bandgap and $G_0W_0$+BSE optical absorption energy. The lowest absorption peak is at 2.44 eV without SOC, 2.42 eV with SOC, thus the exciton binding energy is 0.95 and 0.90 eV, respectively. The SOC effect induces little effect on the binding energy except the splitting bands and doublet emissions as shown in the absorption spectrum (Figure 4b).

Table 1. The lattice constant ($a = b$, Å), bandgap ($E_{PBE}^{K_c-\Gamma_v}$, $E_{PBE}^{K_c-K_v}$, $E_{GW}^{K_c-\Gamma_v}$, $E_{GW}^{K_c-K_v}$), optical gap ($E_{GW+BSE}^{opt}$), the static dielectric constant ($\varepsilon_1$) and exciton binding energy ($E_b^{e-h}$) of monolayer $MoSi_2N_4$ under 0% ~ 6% strain were obtained by PBE, $G_0W_0$ and BSE. All energies are in the unit of eV.

| Strain | $a$ | $E_{PBE}^{K_c-\Gamma_v}$ | $E_{PBE}^{K_c-K_v}$ | $E_{GW}^{K_c-\Gamma_v}$ | $E_{GW}^{K_c-K_v}$ | $E_{GW+BSE}^{opt}$ | $\varepsilon_1$ | $E_b^{e-h}$ |
|---|---|---|---|---|---|---|---|---|
| 0 | 2.91 | 1.73 | 2.04 | 3.19 | 3.39 | 2.44 | 3.90 | 0.95 |
| +SOC | | | | 3.17 | 3.32 | 2.42 | 3.16 | 0.90 |
| 1% | 2.94 | 1.51 | 1.90 | 2.91 | 3.21 | 2.30 | 3.97 | 0.91 |
| 2% | 2.97 | 1.31 | 1.78 | 2.67 | 3.04 | 2.15 | 4.06 | 0.89 |
| 3% | 3.00 | 1.14 | 1.66 | 2.45 | 2.89 | 2.02 | 4.14 | 0.87 |
| 4% | 3.03 | 0.97 | 1.55 | 2.24 | 2.75 | 1.90 | 4.25 | 0.85 |
| 5% | 3.06 | 0.83 | 1.45 | 2.06 | 2.61 | 1.78 | 4.36 | 0.83 |
| 6% | 3.09 | 0.71 | 1.36 | 1.89 | 2.49 | 1.69 | 4.49 | 0.80 |

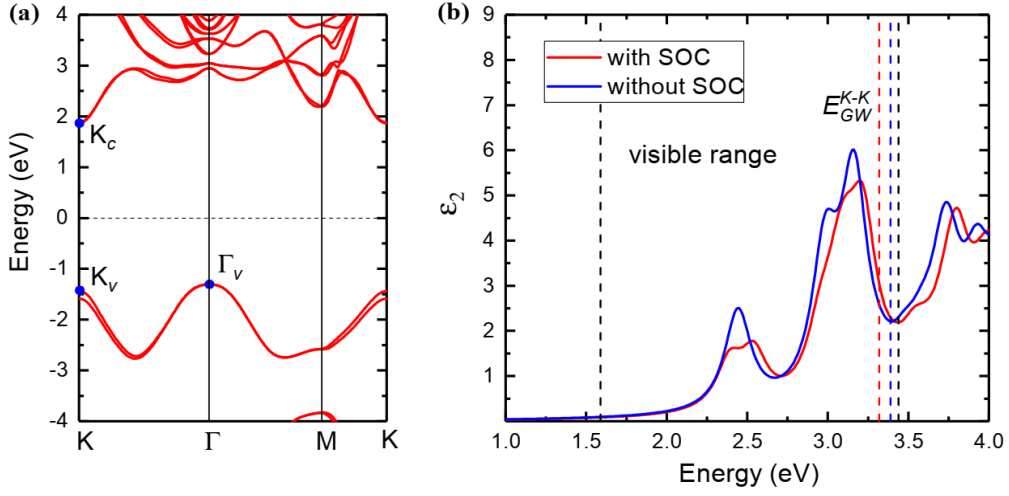

FIG. 4. (a) Band structure of MoSi$_2$N$_4$ using G$_0$W$_0$ with SOC taken into account. (b) Comparison of imaginary part ($\varepsilon_2$) of dielectric functions with and without SOC. The red and blue dashed lines represent minimum direct G$_0$W$_0$ bandgap, two black dashed lines set the visible range.

The G$_0$W$_0$+BSE spectra for MoSi$_2$N$_4$ with tensile strains are plotted in Figure 5a. Considering the small effect of SOC on the binding energy, and to focus on the trends induced by the strains, therefore, for strained MoSi$_2$N$_4$ the SOC is not included. Two bright absorption peaks in optical spectra below the QP direct gap can be identified, the first peak at lowest-energy (1.7 ~ 2.5 eV) arises from direct transition at K-point, which can be labeled 1$s$ exciton ground state. At energies above 1$s$ exciton, one more excited exciton state (2.2 ~ 3.0 eV) labeled 2$s$ is due to the direct transition at Γ-point partially coinciding with other indirect excitons between K- and Γ-point. Notably, higher energy excitonic states tend to be more affected by electron-phonon interactions and are more spectrally broadened (obviously under 0% ~ 2% strain). According to the calculated squares of transition dipole moment from the top band of VB and bottom band of CB, as shown in Figure 2c, the largest dipole transition occurs at K-point, together with weaker but apparent coupling at Γ-point. Obviously, the position of the lowest-energy emission peak in the imaginary part of dielectric function correlates nearly linearly with tensile strains (exact values are given in Table 1), consistent with the reduced electronic bandgap. The coefficient calculated via $dE_g / d\varepsilon$ for calibrating the strain dependent optical bandgap is found to be 0.12 eV, indicating a strong modulation of excitonic emission.

Inferred from the difference between the QP $K_c$-$K_v$ direct gap and BSE optical gap, our calculated results demonstrate that $MoSi_2N_4$ shows a large and steady exciton binding energy which gently reduces from 0.95 to 0.80 eV up to 6% strain, a decrease by 0.02 eV per 1% strain (Figure 5a). Therefore, the electronic and optical gaps of monolayer $MoSi_2N_4$ are controllable by tensile strains, which can be efficiently tuned by strain engineering, whereas the exciton binding energy is relatively stable according to our current results. This seems to be similar to $MoS_2$, in which the exciton binding energy is insensitive to tensile strains [24]. This robust energetics of excitons would be rooted in the well-capped hole and electron frontier states that are mainly contributed by Mo-$d$ states regardless of tensile strains. We plot the real part of dielectric function, absorption and reflection spectra of $MoSi_2N_4$ under 0%, 2%, 4% and 6% strain using $G_0W_0$+BSE, as shown in Figure 5b and c. It is noted that the static dielectric constant (real part of the dielectric function at zero energy) increases as tensile strains, all static dielectric constant values are listed in Table 1. Both $MoSi_2N_4$ and strained $MoSi_2N_4$ exhibits strong absorption in the visible range, as the absorption spectrum shows an obvious redshift with applying tensile strains. Moreover, compared to unstrained $MoSi_2N_4$, more absorption peaks are emerged in the visible rang with tensile strains, accompanying with a decrease in absorption intensity. Similar behavior is observed in the reflection spectra, in which the reflectivity reduces with increasing strains.

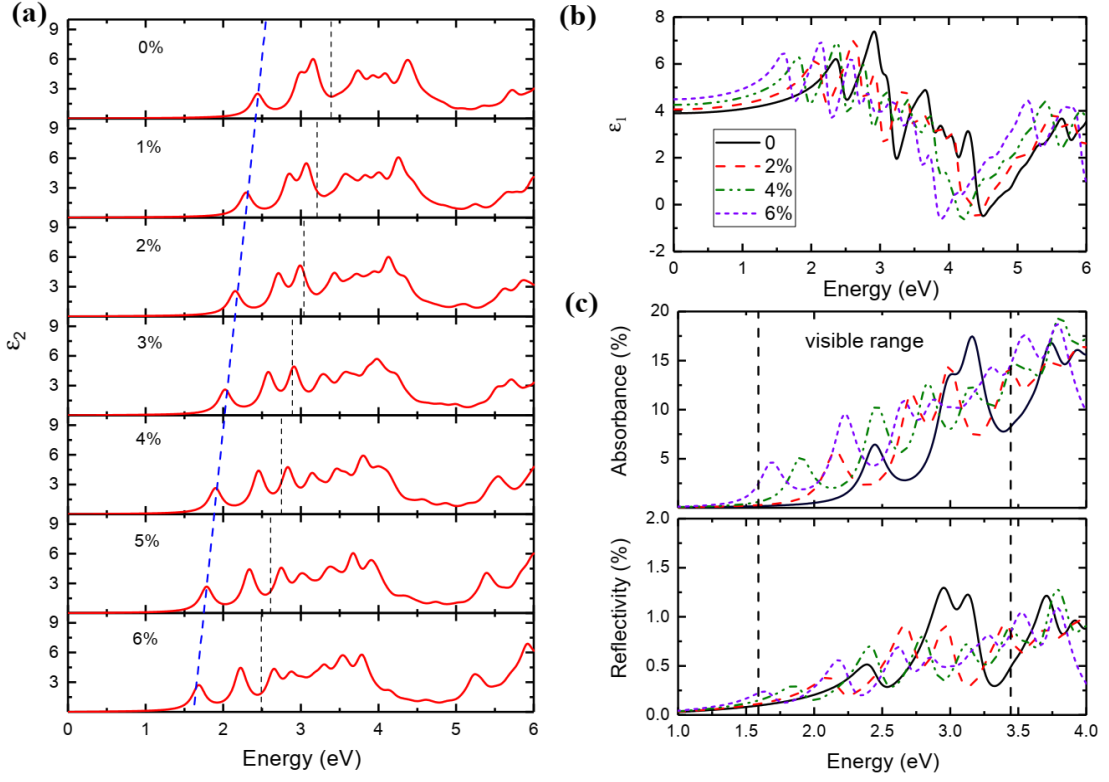

FIG. 5. (a) Imaginary part ($\varepsilon_2$) of dielectric function of unstrained and strained $MoSi_2N_4$. The blue dashed line in (a) indicates the first absorption peak and black dashed line represents the QP $K_c$-$K_v$ direct gap. (b-c) Real part ($\varepsilon_1$) of dielectric function, absorption and reflection spectra of $MoSi_2N_4$ under 0%, 2%, 4% and 6% strain obtained by $G_0W_0$+BSE. The black dashed lines represent the visible range.

## IV. CONCLUSIONS

In summary, we examined the quasiparticle behaviors and optical spectra of monolayer $MoSi_2N_4$ by using GW and BSE methods from first-principles calculations. The excitons in this intrinsically dually novel sandwiched structure are found to be highly localized at the Mo center plane and have a strong binding energy up to 0.95 eV. While the electronic and optical gaps are sensitive to tensile strains, the strong exciton binding energy is largely maintained for lattice stretching up to 6% according to our calculations. Interestingly, the $MoSi_2N_4$ exhibits a strong absorption in the visible range, and the tensile strains causes an obvious redshift of excitonic peaks. Such robust emissions are associated with the well-protected frontier orbitals located in the central $MoN_2$, capped by the surface SiN layer, which leads to the highly confined and bound excitons against environmental disturbances.

Note: During the preparation of the manuscript, we noticed a latest study [34]

showing the quasiparticle and optical properties of $MoSi_2N_4$ can be modulated through the nonlocal dielectric screening effects via controlling thickness of layers.